\documentclass[10pt,conference]{IEEEtran}
\IEEEoverridecommandlockouts
\pdfoutput=1

\usepackage[T1]{fontenc}
\usepackage[utf8]{inputenc}
\usepackage{microtype}
\usepackage{inconsolata} 
\usepackage{amsmath}
\usepackage{graphicx}
\usepackage{latexsym}
\usepackage{indentfirst}
\usepackage{float}
\usepackage{booktabs}
\usepackage{caption}     
\usepackage{lscape}
\usepackage{algorithm}
\usepackage{algorithmicx}
\usepackage[noend]{algpseudocode}
\usepackage{multirow}
\usepackage{balance}
\usepackage{pifont}
\usepackage[table,xcdraw]{xcolor} 
\usepackage{colortbl}
\usepackage{tikz}
\usepackage{enumitem}

\usepackage{cite}

\newcommand{\citet}[1]{\cite{#1}}
\newcommand{\citep}[1]{\cite{#1}}

\newcommand{\circnum}[1]{%
  \ifcase#1 \or \ding{172}\or \ding{173}\or \ding{174}\or \ding{175}\or
  \ding{176}\or \ding{177}\or \ding{178}\or \ding{179}\or \ding{180}\fi
}

\newcommand{\blackcircnum}[1]{%
  \tikz[baseline=(char.base)]{
    \node[shape=circle, fill=black, text=white, inner sep=0.8pt, font=\small] (char) {#1};}%
}

\definecolor{lightblue}{RGB}{235,245,255}

\usepackage[hidelinks]{hyperref}

\AtBeginDocument{}

\title{AgentDroid: A Multi-Agent Tool for Detecting Fraudulent Android Applications}

\author{\IEEEauthorblockN{Ruwei Pan}
\IEEEauthorblockA{panruwei@stu.cqu.edu.cn \\
Chongqing University\\
Chongqing, China \\}
\and
\IEEEauthorblockN{Hongyu Zhang*}
\IEEEauthorblockA{hyzhang@cqu.edu.cn \\
Chongqing University\\
Chongqing, China \\}
\and
\IEEEauthorblockN{Zhonghao Jiang}
\IEEEauthorblockA{zhonghaojiang@cqu.edu.cn \\
Chongqing University\\
Chongqing, China \\}
\and
\IEEEauthorblockN{Ran Hou}
\IEEEauthorblockA{houran@stu.cqu.edu.cn \\
Chongqing University\\
Chongqing, China \\}
}

\begin{document}
\maketitle

\begin{abstract}

\noindent
With the increasing prevalence of fraudulent Android applications such as fake and malicious applications, it is crucial to detect them with high accuracy and adaptability. We present AgentDroid, a novel tool for Android fraudulent application detection based on multi-modal 
analysis and multi-agent systems. AgentDroid overcomes the limitations of traditional detection methods such as the inability to handle multimodal data and high false alarm rates. It processes Android applications and extracts a series of multi-modal data for analysis. Multiple LLM-based agents with specialized roles analyze the relevant data and collaborate to detect complex fraud effectively. We curated a dataset containing various categories of fraudulent applications and legitimate applications and validated our tool on this dataset. Experimental results indicate that our multi-agent tool based on GPT-4o achieves an accuracy of 91.7\% and an F1-Score of 91.68\%, 
outperforming the baseline methods. A video of AgentDroid is available at \url{https://youtu.be/YOM9Ex-nBts}.
\end{abstract}


\section{Introdution}
\noindent
As the Android app market continues to grow, fraudulent apps have become an increasingly significant security threat. 
Various types of fraudulent applications—a subset of malicious applications involving scam, sex, and gambling—can compromise data integrity, disrupt system availability, and leak private data\citep{lu2020android}. 
For instance, research shows that ad fraud alone costs mobile advertisers up to \$1.3 billion in 2015\citep{gersen2016mobile}, and this number has likely increased in recent years. 
Thus, it is critical to develop innovative and adaptive solutions to identify and detect fraudulent Android applications\citep{arp2014drebin}.

A variety of methods \citep{dong2018frauddroid} have been developed for detecting fraudulent applications, including feature extraction and analysis mechanisms. These methods collect characteristics of Android applications through static analysis tools and analyze the extracted data to detect fraudulent applications without executing Android package files. Android Package (APK) files contain multimodal data, including MD5 hashes, network addresses, operating system permissions, icons, and more. Each of these data types can provide crucial insights into fraudulent behavior. For example, fraudulent applications often use generic or low-quality icons that do not align with their advertised functionality or purpose, which can serve as an indicator of fraud.
However, there are various limitations in detecting fraudulent applications using traditional static analysis \citep{arp2014drebin}, such as high false positive rates, poor generalization of machine learning models, and the inability to process multimodal data.

To address these challenges, we develop AgentDroid, a tool for detecting fraudulent applications based on multimodal analysis and a multi-agent system. 
The multi-agent system leverages multimodal analysis to integrate information from multiple data sources (such as fingerprint information and application icons) and combines the collaborative work of multiple agents to achieve a comprehensive understanding of application behavior. Our tool, AgentDroid, assigns agents specialized in analyzing multimodal data. These agents work in parallel and share findings with a Task Master agent, which coordinates their actions and consolidates results.  
In our experiments, we validated AgentDroid using a dataset of over 600 APK files, including both fraudulent and legitimate applications.
The experimental results show that our approach can reduce the false alarm rate and improve detection accuracy.

We summarize our contributions as follows: 
\begin{itemize}
    \item We propose \textbf{AgentDroid}, a multi-agent collaborative tool that leverages \textbf{multi-modal data and specialized LLM-driven agents}, enabling accurate detection of fraudulent Android applications. Unlike traditional single-model approaches, AgentDroid can effectively handle heterogeneous features such as text, icons, and certificates in a unified and interpretable framework.
    \item We evaluate the effectiveness of AgentDroid through experiments, and the results show that our tool outperforms existing methods.
    \item We have open-sourced our tool and dataset at \url{https://github.com/Wwstarry/LLM4Fraud}. 
\end{itemize}

\vspace{-6pt}
\section{Related Work}

\noindent
Early approaches detect Android malware using static features and machine learning algorithms, including Drebin\citep{arp2014drebin},  which extracts permissions, API calls, and other manifest-based features to train a linear SVM classifier for fraudulent detection.
Meanwhile, deep learning-based models have been widely explored. 
For instance, 
Yuan et al. \citep{yuan2016droiddetector} implement an online deep-learning-based Android malware detection engine (DroidDetector) that can automatically detect whether an app is fraudulent or not. Recent methods such as AppPoet \citep{zhao2025apppoet} and LAMD \citep{qian2025lamd} use LLMs to process multimodal app data through prompt-based reasoning. 
AppPoet statically extracts multi-view features of apps and uses prompt-guided LLMs with a DNN classifier to semantically analyze and detect fraudulent Android applications, and LAMD is a context-driven framework that enables effective Android fraudulent detection by guiding LLMs through tiered reasoning over security-critical code regions.
Unlike existing models, our tool adopts a collaborative multi-agent design, enabling specialized agents to reason over different data types and improve interpretability and adaptability.




\section{Proposed Approach}

\begin{figure*}[ht]
\centering
\includegraphics[width=\textwidth]{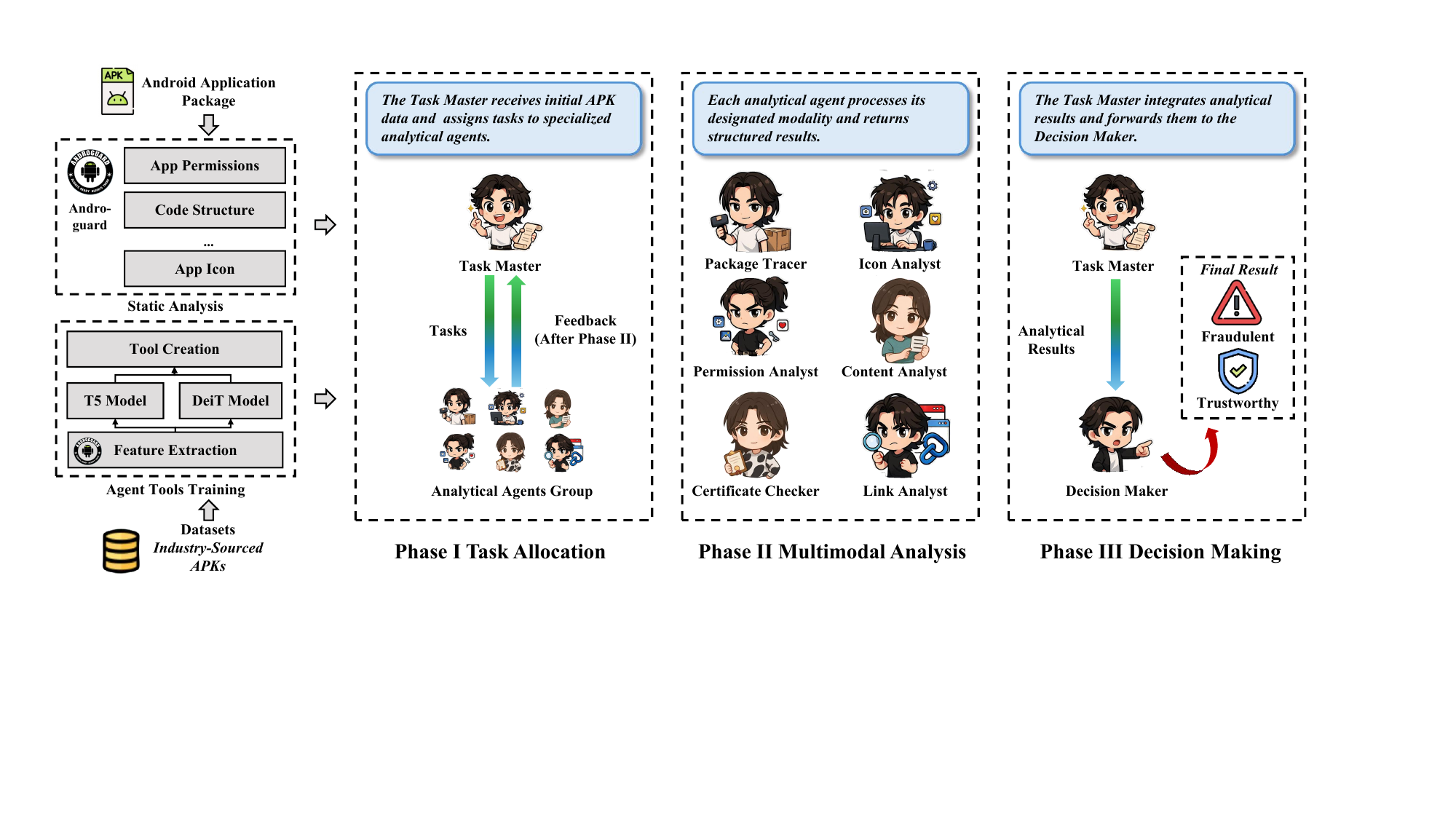} 
\caption{An Overview of AgentDroid}
\label{fig:overview}
\end{figure*}

\noindent
We propose AgentDroid, a multi-agent fraud detection tool that uses static analysis and multimodal features extracted from APK files. Figure \ref{fig:overview} provides an overview of AgentDroid. 
The tool analyzes metadata, permissions, icons and source code to determine the presence of fraudulent behavior and classify the app into specific fraud categories such as gambling and so on. 

\subsection{Static Analysis and Agent Tools}
\noindent
AgentDroid performs static analysis on each APK using Androguard \citep{desnos2018androguard}, extracting multimodal data such as metadata, permissions, icons, and decompiled source code. 
Specifically, we analyze the DEX bytecode, the AndroidManifest.xml, the resource files, and so on.
These features are then processed by specialized agents to detect whether the app is fraudulent.





To support this analysis, AgentDroid provides agents with querying interfaces for interpreting metadata, image classifiers (e.g., DeiT model \citep{touvron2021trainingdataefficientimagetransformers}) for visual features (icons), and text classifiers (e.g., T5 model \citep{raffel2020exploring}) for textual features (permissions, descriptions) to inspect for fraudulent patterns. Source code is made accessible via a structured interface, allowing agents to search, navigate, and interpret program logic. Each agent is equipped with domain-specific tools and collaborates within the multi-agent system to contribute findings for the final decision-making.

\subsection{Phase I Task Allocation}

After Static Analysis and Agent Tools Training, the agents collaborate according to a pre-designed task allocation strategy, described in Algorithm \ref{alg}. Firstly, the Task Master receives initial APK features and assigns tasks to specialized analytical agents.
Each agent is responsible for a specific task and can utilize specific tools to perform the required analysis.
The list of agents in our approach, all powered by LLMs, is as follows:
\textbf{\blackcircnum{1} Task Master}: Assigns tasks to the appropriate agents and integrates their results for decision‐making.  
\textbf{\blackcircnum{2} Package Tracer}: Retrieves the app package information, including the app name, activities, and related metadata.  
\textbf{\blackcircnum{3} Icon Analyst}: Analyzes the app’s icon to detect if it resembles icons from known fraudulent apps.  
\textbf{\blackcircnum{4} Permission Analyst}: Analyzes the app’s requested permissions to assess potential risks or sensitive behaviors.  
\textbf{\blackcircnum{5} Content Analyst}: Analyzes textual content within the app for any suspicious or illicit activity indicators.  
\textbf{\blackcircnum{6} Certificate Checker}: Verifies the authenticity and validity of the app’s certificate to identify possible fraud.  
\textbf{\blackcircnum{7} Link Analyst}: Inspects relationships between the given app and other related apps to uncover hidden connections.  
\textbf{\blackcircnum{8} Decision Maker}: Aggregates all results and provides the final decision on whether the app is fraudulent.  

Although some agents, such as the Permission Analyst and Content Analyst, both operate on textual inputs, we design them as separate modules due to their distinct data modalities and domain-specific reasoning requirements. 
Permission data is structured and typically requires an understanding of Android API semantics, while content analysis involves unstructured, natural-language understanding of user-facing elements such as descriptions or messages. 
Keeping these roles separate allows each agent to employ optimized tools (e.g., fine-tuned T5 for structured text) and improves adaptability to evolving fraud patterns.

The eight agents are broadly divided into two categories: one responsible for decision-making and the other responsible for analysis. As illustrated in Figure \ref{fig:overview}, the decision-making agents include the Task Master and the Decision Maker, while the analytical agents include the Certificate Checker, the Package Tracker, the Link Analyst, the Permission Analyst, the Icon Analyst, and the Content Analyst.
All the agents in AgentDroid are powered by the GPT-4o model by default,
which allows them to perform advanced reasoning and collaboration.

\begin{algorithm}[t]
\caption{Dynamic Task Allocation by Task Master}
\begin{algorithmic}[1]
\Require Initial data $collectedData$
\Ensure Final result $finalResult$
\State Initialize $tasksToAssign \gets$ Determine initial tasks based on $collectedData$
\State Initialize $collectedResults \gets$ empty set
\While{cannot-make-decision($collectResults$)}
    \For{each $task$ in $tasksToAssign$}
        \State $agent \gets$ Select the corresponding agent($task$)
        \State Assign $task$ to $agent$
        \State $result \gets$ Receive result from $agent$
        \State Add $result$ to $collectedResults$
    \EndFor
    \State Integrate $collectedResults$ and update $collectedData$
    \State $tasksToAssign \gets$ Analyze and determine new tasks based on the updated $collectedData$
\EndWhile
\State Integrate all $collectedResults$
\State Pass the integrated information to the final agent $finalAgent$
\State $finalResult \gets$ Receive final result from $finalAgent$
\State \Return $finalResult$
\end{algorithmic}
\label{alg}
\end{algorithm}

\subsection{Phase II Multimodal Analysis}

\begin{figure}[t]
    \centering
    \includegraphics[width=0.5\textwidth]{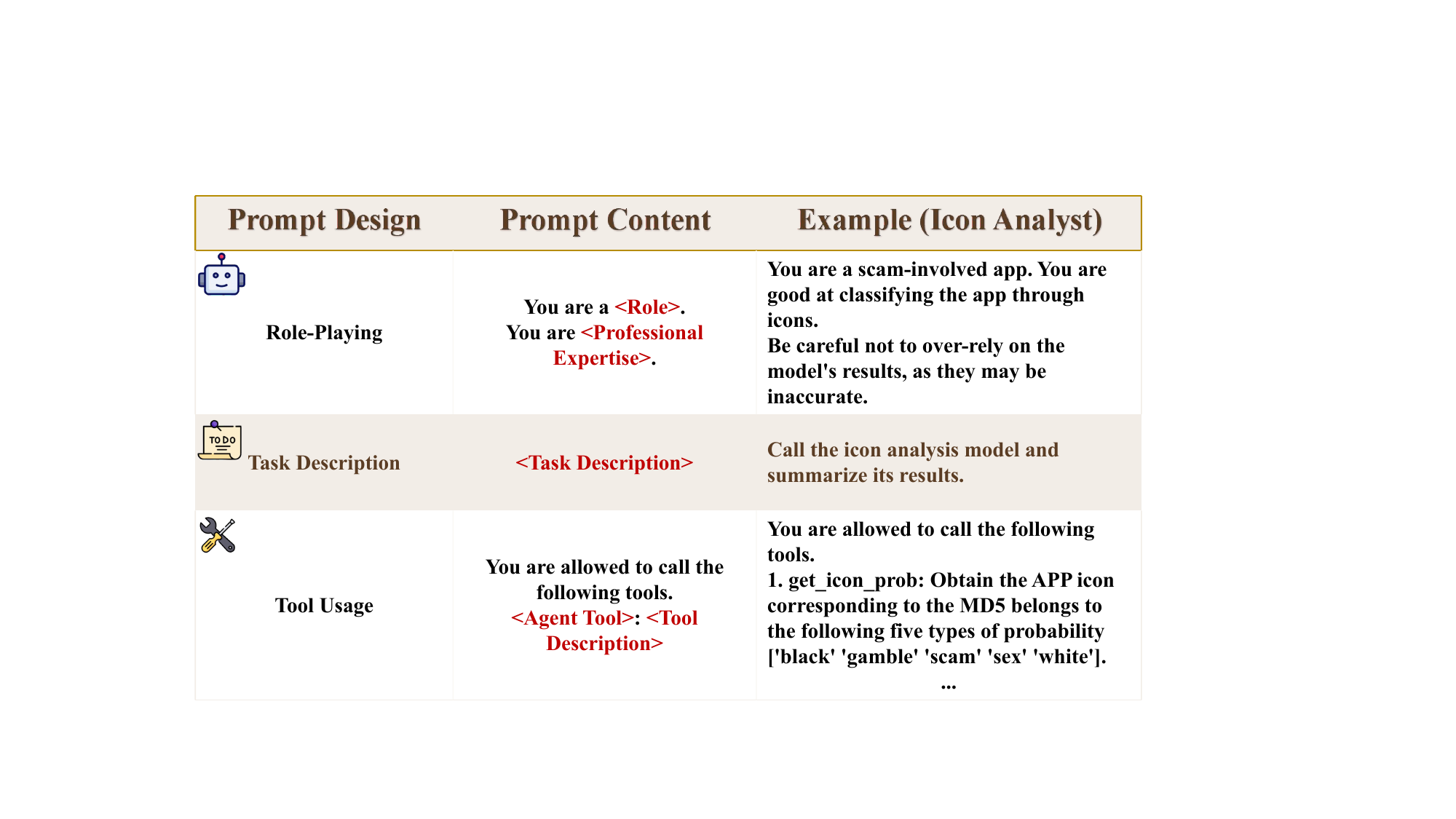}
    \caption{Prompt Template used by AgentDroid. The complete prompt template is available at \href{https://github.com/Wwstarry/LLM4Fraud}{our repository}.}
    \label{fig:prompt}
\end{figure}

In the Multimodal Analysis, which is the core phase of our tool consisting of specialized agents collaborating for fraud detection, we use LangGraph and Prompt Engineering to create eight specialized agents responsible for task scheduling, icon analysis, content analysis, certificate inspection, package tracking, sensitive permission analysis, relationship analysis, and final decision-making.
An example of the prompt design used to guide the agents is shown in Figure \ref{fig:prompt}, which illustrates the role-playing structure, task description, and allowed tools for the Icon Analyst.
Each analytical agent, such as the Package Tracker or Icon Analyst, specializes in processing specific types of data, such as certificates, permissions, icons, and content. 
Upon completing their tasks, the analytical agents return their results to the Task Master for integration.
The Task Master orchestrates the process by determining whether additional information is needed based on collected data and assigns specific tasks to the analytical agents. 
The collaboration process among agents is coordinated by the Task Master, which maintains a shared knowledge state (\texttt{collectedData}) that aggregates intermediate results. When an agent completes a task, it returns structured output back to the Task Master. This output is appended to the shared state and can trigger new tasks that depend on the updated context. For instance, if the Package Tracer detects suspicious app naming patterns, this information may prompt further investigation from the Link Analyst. All agents communicate asynchronously via the LangGraph backend, and intermediate states are stored in memory to support iterative reasoning.


\subsection{Phase III Decision Making}
In this phase, the Task Master 
passes the consolidated information to the Decision Maker, who evaluates the aggregated data to determine the final category and probability of fraud for the APK. 
The Task Master coordinates agent execution based on task dependencies and data availability. For instance, if an app lacks certificate information, the Certificate Checker is skipped to prevent unnecessary analysis. Likewise, if the Permission Analyst detects suspicious access (e.g., to SMS or contacts), the Task Master can prioritize content inspection by the Content Analyst. This design enables adaptive task sequencing and avoids redundant or irrelevant processing.
The eight agents are designed based on common fraud behaviors, such as fake certificates, icon impersonation, permission abuse, and scam-related content. Each agent targets one specific fraud type to ensure clearer analysis. The tool is modular and supports adding new agents to detect emerging fraud patterns.

\section{Evaluation}

\subsection{Experimental Setup}

\begin{figure}[t]
    \centering
    \includegraphics[width=0.5\textwidth]{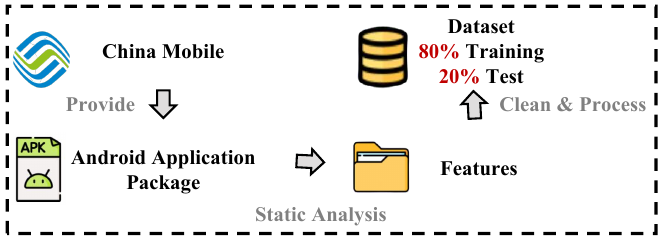}
    \caption{The workflow of curating the dataset.}
    \label{fig:dataset}
\end{figure}

\noindent
We curated a dataset consisting of 660 APKs, encompassing 480 fraudulent applications and 180 legitimate applications, shown in Figure \ref{fig:dataset}. 
This dataset is designed to train and validate the performance of our multi-agent system in detecting Android fraudulent applications. 
The APK files were from a major telecommunications enterprise(China Mobile). 
Fraudulent labels were officially assigned by the enterprise based on real-world network activity, making the dataset both representative and reliable for fraud detection research.
The original data is available at: \url{https://github.com/Wwstarry/LLM4Fraud}
Before conducting the experiments, we performed a static analysis on the APKs to extract key features including metadata, permissions, and icons. The extracted features were then cleaned and processed, after which the dataset was split into training (80\%) and test (20\%) sets to fine-tune and evaluate the performance of the tool across various modalities, including text and icon classification.

\begin{table}[h!]
\centering
\caption{Comparison of Various Models for Fraudulent Android App Detection}
\resizebox{\columnwidth}{!}{
\begin{tabular}{lcccc}
\hline
\textbf{Model} & \textbf{ACC(\%)} & \textbf{Precision(\%)} & \textbf{Recall(\%)} & \textbf{F1(\%)} \\
\hline

BERT         & 44.53                &46.19                  & 44.53               &45.34               \\
Clip      & 33.33                & 31.50                & 33.33               & 32.39               \\
Resnet-18  \citep{he2016deep}      & 47.62              & 47.11              & 47.62               & 47.36                 \\
Drebin \citep{arp2014drebin}         & 40.15                  & 44.20                   & 40.15                & 42.08                   \\
DroidDetector \citep{yuan2016droiddetector} & 61.63  &  61.80  & 47.83   & 53.92\\

GPT-4o & 78.09                & 79.91          & 81.02          & 80.46               \\
AppPoet \citep{zhao2025apppoet} & 82.58  & 89.35   & 76.89 & 82.66 \\
LAMD \citep{qian2025lamd} & 88.34   &  88.7   & 83.61 & 86.08   \\
\rowcolor{lightblue}
AgentDroid (GPT-4o)& \textbf{91.70 }                 & \textbf{91.79    }               & \textbf{91.70 }               & \textbf{91.68   }                \\

\hline
\end{tabular}
}
\label{tab: result}
\vspace{-6pt}
\end{table}


In the experiments, all the LLMs in AgentDroid and baselines are powered by the GPT-4o model with a temperature of 0.5, which provides fast and accurate reasoning at a lower cost.
We compare AgentDroid against several representative baselines, including machine learning, deep learning, and LLM methods. 
To ensure robustness, each experiment is repeated multiple times and the average results are reported.
All methods were fine-tuned using the same 80\% training split to ensure fair comparison.
For deep learning, we adopt ResNet-18 for icon classification and BERT for permission text classification. CLIP, a multimodal vision-language model, is used to jointly encode app icons and textual descriptions for fraud prediction.
To benchmark the effectiveness of the multi-agent design, we also compare against a single-agent GPT-4o baseline, in which GPT-4o acts as a zero-shot classifier and directly performs inference on the 20\% test set without training.
We adopt four key metrics: Accuracy (ACC for short), Precision, Recall, and F1-Score for the evaluation.



\subsection{Results}

\noindent
As shown in Table \ref{tab: result}, AgentDroid achieves the highest performance across all evaluation metrics, with an Accuracy of 91.70\% and an F1 score of 91.68\%. Compared to traditional models such as Drebin and DroidDetector, as well as recent LLM-based models including AppPoet and LAMD, our tool demonstrates substantial improvement. 
The inferior performance of single-modality baselines such as BERT (text-only) and ResNet-18 (icon-only) highlights the necessity of integrating multimodal features.
Notably, GPT-4o used in a single-agent zero-shot setting performs poorly with an F1 score of 80.46\%, underscoring the effectiveness of our multi-agent collaborative design. These results confirm the advantage of integrating multimodal analysis with agent specialization for robust fraud detection.

\section{Conclusion and Future Work}
\noindent
In this paper, we propose an Android fraudulent application detection tool named AgentDroid, which is based on multimodal analysis and multi-agent collaboration.
Our preliminary evaluation demonstrates the effectiveness
of AgentDroid.
Through multi-agent collaboration and the use of specialized agent tools, our approach achieves higher detection accuracy than existing methods, demonstrating its potential practicality. The dataset and our source code are publicly available at \url{https://github.com/Wwstarry/LLM4Fraud}.

In future work, we plan to extend AgentDroid to incorporate dynamic analysis techniques to capture runtime behaviors of applications.
We also intend to evaluate the tool on larger and more diverse datasets to further validate its generalizability and robustness.




\bibliographystyle{IEEEtran}
\bibliography{references_acl} 

\end{document}